# Analysis of visitors' mobility patterns through random walk in the Louvre museum

Yuji Yoshimura[a], Roberta Sinatra[b,c,d], Anne Krebs[e], Carlo Ratti[a]

[a] SENSEable City Laboratory, Massachusetts Institute of Technology, 77 Massachusetts Avenue, Cambridge, MA 02139, USA;

[b] Department of Mathematics and its applications, and Department of Data and Network Science, Central European University, Nador utca 9, Budapest 1067, Hungary;

[c] Network Science Institute, Northeastern University, 177 Huntington Avenue, Boston, MA 02116, USA;

[d] Complexity Science Hub, Josefstaedter Strasse 39, 1080 Vienna, Austria;

[e] Dominique-Vivant Denon Research Centre, musée du Louvre, 75058, Paris, Cedex 01, France;

**Abstract.** This paper proposes a random walk model to analyze visitors' mobility patterns in a large museum. Visitors' available time makes their visiting styles different, resulting in dissimilarity in the order and number of visited places and in path sequence length. We analyze all this by comparing a simulation model and observed data, which provide us the strength of the visitors' mobility patterns. The obtained results indicate that shorter stay-type visitors exhibit stronger patterns than those with the longer stay-type, confirming that the former are more selective than the latter in terms of their visitation type.

## 1. Introduction

This paper analyzes visitors' mobility patterns in the Louvre, a large museum in Paris. We deployed seven Bluetooth sensors across the museum, which provided large datasets of visitors' behaviors. The question asked is whether or not we can measure the strength of their patterns and evaluate them in the same framework, independent of their path sequence length. Because the probability of a shorter path sequence's appearance is much higher than that of a longer path sequence in nature, we propose a random walk model in a large-scale network and compare observed and simulated data.

In art museums and galleries, the systematic observation of visitors' movements has been conducted in the form of "timing and tracking", most likely relying on the paper-and-pencil method (Yalowitz and Bronnenkant, 2009, p. 52). These quantitative methods were used in the early visitor studies (Robinson, 1928; Melton, 1935; Loomis, 1987, pp. 21-23; Hein, 1998, pp. 47-51), based on "the assumption (prevalent at the time) that observations are more objective and reliable than what people might say in interview" (Hooper-Greenhill, 2006, p. 363). On the contrary, Wittlin (1949) used the qualitative method of collecting visitors' behavior "by asking people open-ended questions and interpreting their answers as well as their drawings" (Hein, 1998, p. 51). Although Hein pointed out that the characteristics of a series of Robinson and Melton's works were "the large sample sizes - timing studies of adults used samples up to almost 2,000" (1998, p. 49), most of the datasets concerning art museums and galleries contain relatively small samples gathered from a spatially confined area, even those that use new and emerging technology.

**Table 1.** Data collection techniques to study visitors' paths in the context of museums

|  | **Data capture** | **Obtained data** | **Sample number** | **Result** |
| --- | --- | --- | --- | --- |
| Melton (1935) | Observations | Paths through a gallery, the length of stay in each room, the spent time looking at each exhibit, number of stops by visitors | Almost 2,000 | Turning right at choice points in galleries, attracting-power, holding-power (see Loomis, 1987, page 21-23 for a summary) |
| Serrell (1998) | Observations | The length of stay in the exhibition, the number of stops at the exhibits | 8,507 from 110 exhibitions (max: 458) | Average length of stay divided into exhibition's square footage (SRI), visitors who stop at more than half the exhibits (DVI) |
| Bourdeau and Chebat (2001) | Observations and questionnaire | Visitors' movements, sketches of the pathways made by each visitor as reminiscences | 60 | The influence of the design of display to visitor flow |
| Sparacino (2002) | Infrared location sensors | Path and length of stops | 50 | Visitors' types; Busy, Selective, Greedy |
| Kanda et al, (2007) | RFID | Sequential movement between sensors, and the length of stay around the sensors | 8,091 | Typical spatial use and visiting pattern, Spatial division by usage of visitors |
| Tschacher et al, (2012) | Wearable gloves and questionnaire, tag with ultrawide band signals of 6-8 GHz | Locomotion, heart rate and skin conductance, visitors' paths with a precision of 15cm | 373 | Relationship between physiological responses and aesthetic-emotional experiencing in the exhibits |
| Yoshimura et al, (2012) | Bluetooth Sensor | Sequential movement between exhibits where sensors were installed, length of stay in the museum | 12,944 | Spatial usage and visitors' behaviors based on their trajectories and length of stay |

Table 1 shows an example of the data collection technique for the paths of visitors in a museum, the sample size of the obtained data, and the results of the analysis using these datasets. One of the most relevant tools for recording visitors' behavior is optical technology (i.e., time-lapse photographs, video cameras) because it "corroborated subjective impressions" (Nielsen, 1942, p. 109). Loomis described examples of the use of a video camera to analyze visitors' behavior in galleries (Loomis, 1987, p. 221). Alternatively, Sparacino (2002) proposed the use of wearable sensors to collect visitors' behavior in a science museum and distributed the collected samples into previously established categories (i.e., the busy, selective, and greedy) depending on visitors' paths and the length of their stops. Conversely, Kanda et al. (2008) took a bottom-up approach and classified visitors' behavior to analyze their visiting patterns based on the large samples collected by asking visitors to equip RFID tags during their visits to a science museum. Both of these methods were applied in science museums, where much of the previous work has taken place. Art museums and galleries, in contrast, have served as the environmental context in only a few previous studies (Hooper-Greenhill, 2006, p. 368). For those reasons, this paper attempts to collect large datasets in a large art museum.

1.1 Data collection

For this purpose, we employ the Bluetooth detection technique. Bluetooth detection is based on systematic observation that discovers Bluetooth activated mobile devices, putting it in the framework of "unobtrusive measures" (Webb et al., 2000). This makes use of visitors' digital footprints or "data exhaust" (Mayer-Schönberger and Cukier, 2013, p. 113). This technique has been used to collect pedestrians' sequential movements outdoors (Eagle and Pentland, 2005; Paulos and Goodman, 2004; Kostakos et al., 2010; Versichele et al., 2010) as well as indoors (Delafontaine et al., 2012; Yoshimura et al., 2014, 2017). The detection system works as follows: When Bluetooth activated mobile devices enter a sensor's detection range, the sensor keeps detecting their presence until they exit. As each mobile device's MAC address is unique in most cases, the sensor network can identify the check-in and check-out of each mobile device with a time stamp. The hash algorithm (Stallings, 2011, pp. 342-361) was applied to maintain the anonymity of visitors' data by converting the MACID into a unique identifier (Sanfeliu et al., 2010). Thus, we collected data about visitors' sequential movements, lengths of stay, and transition times between places while overcoming the privacy issue.

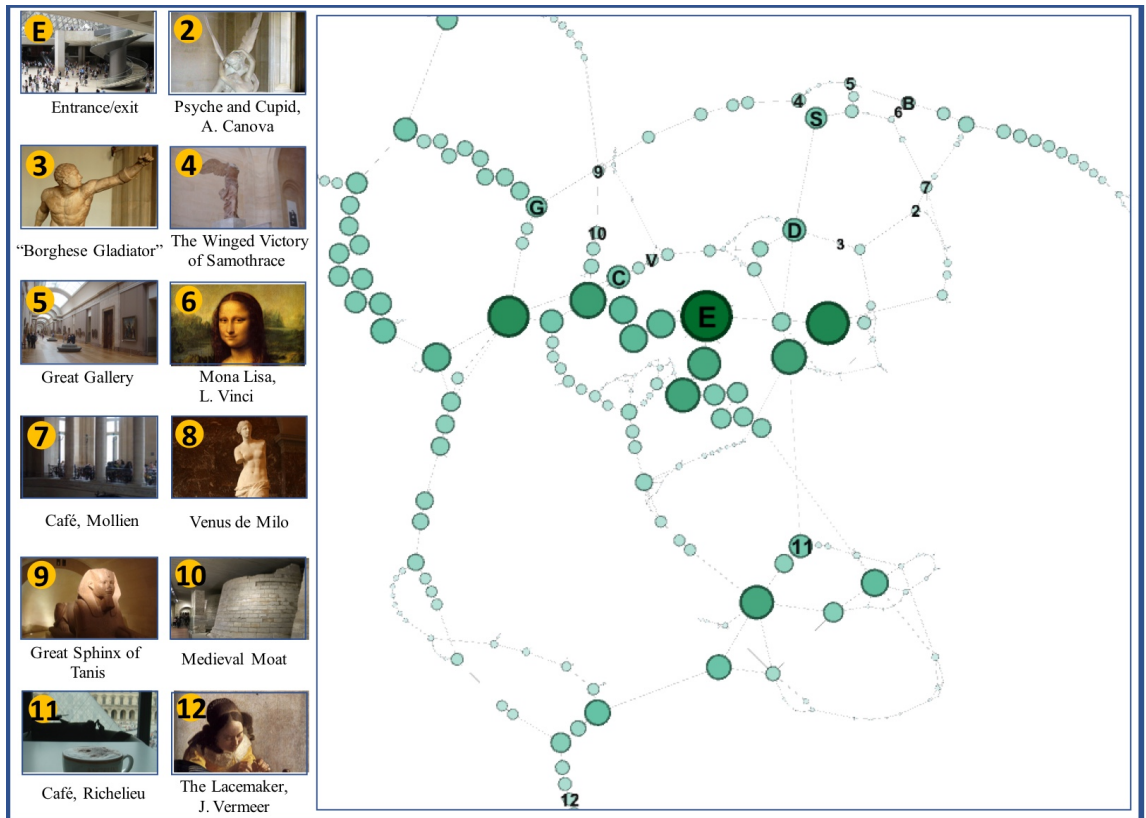

**Figure 1.** The topological representation of the spatial structure of the Louvre Museum. Each node corresponds to a room in the museum, and a link between two nodes corresponds to a corridor to another kind of connection between rooms. The nodes identified by the seven letters correspond to the rooms where sensors are located. The nodes identified by the numbers indicate popular artworks near each sensor. The bigger the circle, the higher the betweenness centrality of the node (see Boccaletti et al. (2006) for a review of betweenness).

Figure 1 presents the topological representation of the spatial structure of the Louvre Museum with the location of Bluetooth sensors and the popular artworks near each sensor. The letters indicate the locations of sensors: Hall (E), Gallery Daru (D), Venus de Milo (V), Salle des Caryatides (C), Great Gallery (B), Victory of Samothrace (S) and Salle des Verres (G). We collected data on 24 days in three different audits: from 30 April to 9 May 2010, from 30 June to 8 July 2010, and from 7 August to 18 August 2010. Within the collected data, we selected the ones that started and finished at node E for the subsequent analysis in order to measure the length of stay in the museum. After data cleanup and processing to remove inconsistencies, 24,452 unique devices were identified and used in our analysis.

## 2. Methodology

### 2-1. Validity of the path length and random walk

The path followed by a visitor through a museum can be represented as a sequence of letters, each of them corresponding to one of the unique locations of the museum. The sequences in the present research are composed of seven letters, i.e., the number of sensors installed in the museum, and start and end with the letter E, which corresponds to the entrance/exit of the museum. Each sequence is at least three letters long, as we only consider visitors who were detected in proximity of one sensor besides the one at the entrance/exit. So, for example, the shortest sequence we observe is of the kind E-S-E. There is in theory no upper bound to the maximum length of sequences, since a person can visit a location more than once. In our study, however, the longest sequence observed consisted of 30 letters.

In general, if we have *N* nodes - in our case 7, the number of sensors installed in the museum - we can compose $N^i$ different sequences of length *i*. In our case, *i* can range from 3, the shortest sequence possible, to 30, the longest sequence observed. So, all possible sequences with a length between these two extremes sum up to

$$\sum_{i=3}^{x} N^x = \frac{N^{x+1} - N^3}{N - 1} \qquad (1)$$

where *x*=30, or the maximum length considered (Sinatra et al., 2010). However, not all the possible sequences are valid for the analysis because there are not enough data to validate long paths (Sinatra et al., 2010). That is, for high values of *x*, we do not have enough paths of length *x* in our observed data to assess the statistical significance of the path. The maximum length *S* of the path that is valid for the analysis can be calculated based on the number of nodes *N* and the sample size of paths (24,452 in our dataset). As we consider only paths starting at the entrance (E) and finishing at the exit (E), we calculate the maximum path length for our analysis by subtracting the fixed first and last nodes; hence, we use S-2. In addition, we consider that the same node is not repeated consecutively. Therefore, the number of possible nodes that can be selected at each step of the visits is *N*-2 (excluding E and the letter of the currently visited node). Taken together, we need to solve the following inequality to determine *S*:

$$(N - 2)^{S-2} < 24{,}452 \qquad (2)$$

which gives *S=8*, meaning that we cannot analyze paths longer than 8.

Moreover, shorter path sequences (i.e., E-S-E) tend to appear more frequently than longer path sequences (i.e., E-D-S-B-D-V-E). Because the probability of the appearance of short combinations of nodes is higher than that of long combinations of nodes, we cannot compare the frequencies of paths of different lengths. In order to solve this problem, we employ a random walk on the graph of *N* nodes to make a reference line for each path length. A random walk on a network is a process simulating walkers moving between nodes in a discrete time. At each timestep, a walker chooses the next node to hop on randomly among those linked to the node the walker is sitting on (Sinatra, et al., 2011). Hence, each movement depends only on the connections of the current location and not on previously visited nodes.

**Figure 2.** Top 20 of ranking of the random walk frequency in case of the longer stay-type visitors with path length less than 7, which is valid for the analysis

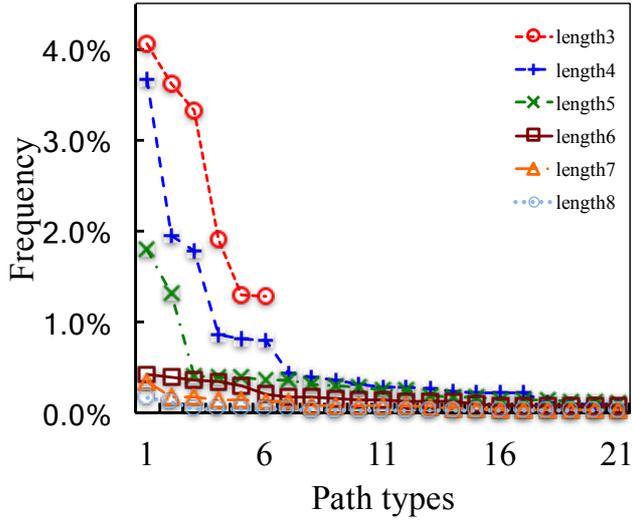

**Table 2.** Top three paths appearing by random walk for longer and shorter stay-type visitors and the respective distributions for each.

| Path | Percentage | Path | Percentage |
|---|---|---|---|
| *Longer Stay Type Visitors* | | *Shorter Stay Type Visitors* | |
| **Path length 3** | | | |
| E-D-E | 4.06% | E-D-E | 6.86% |
| E-S-E | 3.63% | E-C-E | 5.25% |
| E-C-E | 3.33% | E-S-E | 4.56% |
| **Path length 4** | | | |
| E-D-S-E | 3.67% | E-D-S-E | 4.64% |
| E-V-C-E | 1.95% | E-V-C-E | 3.35% |
| E-S-B-E | 1.78% | E-S-B-E | 3.11% |
| **Path length 5** | | | |
| E-D-S-B-E | 1.80% | E-D-S-B-E | 3.17% |
| E-D-V-C-E | 1.31% | E-D-V-C-E | 3.15% |
| E-S-V-C-E | 0.42% | E-S-B-D-E | 0.96% |
| **Path length 6** | | | |
| E-D-S-V-C-E | 0.42% | E-D-S-B-D-E | 0.98% |
| E-D-S-B-D-E | 0.40% | E-S-B-D-S-E | 0.65% |
| E-D-S-D-S-E | 0.36% | E-D-S-V-C-E | 0.55% |
| **Path length 7** | | | |
| E-D-S-B-D-S-E | 0.36% | E-D-S-B-D-S-E | 0.66% |
| E-D-S-D-S-B-E | 0.18% | E-S-B-D-S-B-E | 0.44% |
| E-S-B-D-S-B-E | 0.17% | E-S-B-D-V-C-E | 0.44% |

Table 2 shows the result of the random walk frequency applied to seven nodes. The *x*-axis in Figure 2 takes the rank of the frequently appearing path types at each path length, ordered by number of occurrences. As we can observe, the frequency of each rank in each path length decreases when the path length increases. This reveals the effect, as we mentioned previously, that shorter path lengths appear more frequently than longer path lengths, preventing us from comparing all extracted paths under the same circumstance.

Although a visitor's choice of movements in a museum likely depends on the past locations visited as well as the prior knowledge of the museum (i.e., repeaters or first time visitors), a random walk is a minimalistic model providing a reference line for the frequency of sequences simply induced by the graph structure of the museum. The random walk simulations can therefore provide us with the probability of the transitions between nodes and hence with the probability of each path for a given length. Since the probability of a given path decreases when the path length increases, we cannot compare paths of different lengths with each other. For this reason, we compare the probability of each observed path length with the probability provided by the random walk model. We introduce the *R*-value, defined as the observed frequency of a path divided by the frequency of the same path in the random walk model, in order to measure the strength of each path and identify patterns in the museum. That is, if *R* is large (>>1), the corresponding path is a strong pattern, as it suggests that the observed path appears much more frequently than in the random walk model.

## 2-2. Spearman's correlation

To clarify the distribution of each path, we focus on looking for a correlation between the duration of a visitor's stay in the museum and his or her chosen trajectory. We split our sample into deciles based on the duration of stay, effectively obtaining 10 equally-sized clusters of ~2446 visits each. Meanwhile, we calculate the number of visitors using every possible trajectory across the museum that is present in each of the aforementioned partitions. Then, for every different path, we calculate the correlation between the partition order and the number of visitors in that partition. Since we know that the partitions increase monotonically by duration of stay, if a correlation exists, it can tell us which trajectories are more frequent when a visitor stays for different lengths of time. We quantify the correlation by means of Spearman's correlation coefficient (Corder and Foreman, 2009), denoted by the symbol ρ, and p-value. Spearman's correlation coefficient is used to assess how well the relationship between the two variables (*x*, *y*) can be described using a monotonic function, where the derivative of *y* with respect to *x* is greater than zero, $\frac{dy}{dx}$ > 0 (see appendix).

**Figure 3.** Stay length vs frequency of appearance of four different types of paths normalized by the total number of appearances of each path

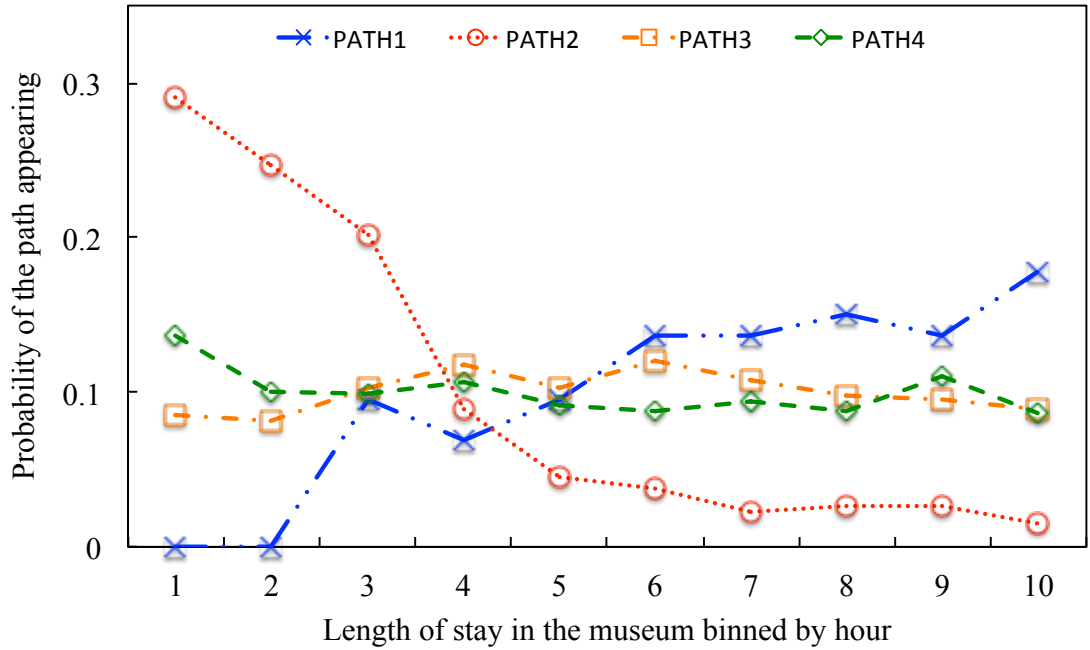

| Path | ρ | p | middev |
| --- | --- | --- | --- |
| PATH1 (E-D-S-B-S-D-V-E) | 0.93% | 5.38E-05 | 0.53% |
| PATH2 (E-S-D-V-C-E) | 0.96% | 1.02E-05 | -3.03% |
| PATH3 (E-S-E) | -0.12% | 0.7328 | 0.53% |
| PATH4 (E-D-S-B-E) | -0.56% | 0.0957 | -1.96% |

To understand the procedure we are using, one can observe Figure 3. All the paths of our datasets can be classified into four types that express their frequencies across the 10 partitions. We can see that the frequency of PATH1 increases as the length of stay increases. Conversely, the frequency of PATH2 decreases as the length of stay increases. According to Spearman's correlation coefficient, PATH1 displays a significant positive correlation with partition order (ρ=0.93, p=5.83E-05), while PATH2 shows a significant negative correlation (ρ=-0.96, p=1.02E-05). This would imply that visitors tend to use PATH1 more when they stay inside the museum longer, thus making PATH1 characteristic of longer stay-type visitors, while PATH2 is chosen more frequently with shorter stay-type visitors. Figure 3 also shows how PATH3 and PATH4 are more evenly distributed across partitions, suggesting that there is no clear tendency to choose any of these paths depending on a visitor's length of stay in the museum.

In addition to this, if we focus on the first partition (equivalent to the shorter stay-type visitors, with a stay length of less than 1:30:27) and the last partition (longer stay-type visitors, with a stay time of more than 05:04:11), we then find trajectories that appear in one of these two partitions but not both, effectively finding frequent paths that are exclusive to both groups. We also quantify correlations of the mean of the p-values for every distinct path in our dataset and their frequencies throughout the dataset.

**3 Results**

## 3.1 Patterns of visitors' sequential movements

This section analyzes the relationship between the distribution of visitors' path types and their frequencies. We extract all path types that appeared in each group and count the number of appearances of each path, resulting in the frequency of each path. As a result, 1,312 different path types are subject to further analysis for the longer stay-type visitors, and 373 for the shorter stay-type visitors. In a similar way, we examine the random walker dataset: 833 for the longer stay-type random walkers and 518 for the shorter stay-type random walkers.

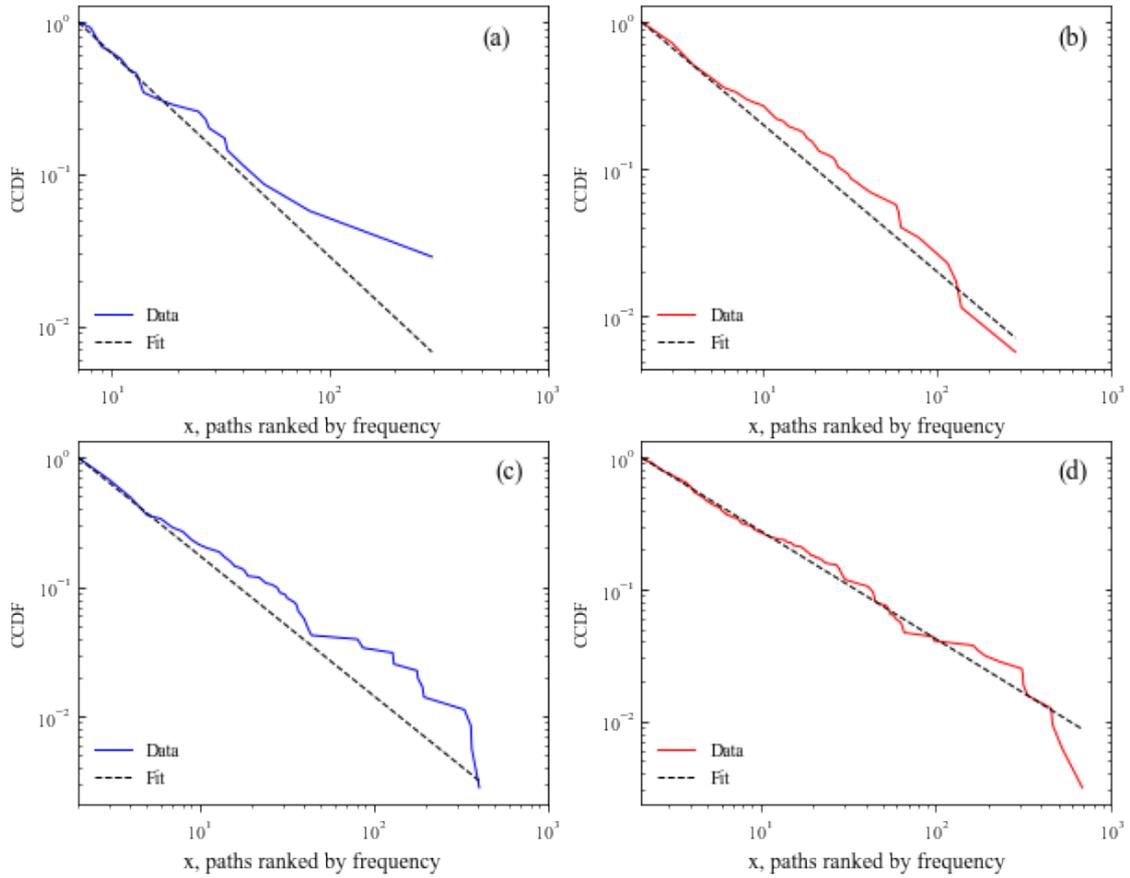

**Figure 4.** (a) Complementary cumulative distribution function (CCDF) of visitors' path types and their frequencies from the longer stay-type visitors. (b) from the shorter stay-type visitors. (c) from the longer stay-type random walkers, and (d) from the shorter stay-type random walkers. The dashed lines in the log-log plots of all panels indicate that p(X>x) follows a power law where the alpha parameter = 2.33, 1.99, 2.08, 1.82 for (a), (b), (c), (d) respectively.

Figure 4 presents the complementary cumulative distribution function (CCDF) of the frequencies of visitors' path types from (a) longer stay-type visitors, (b) shorter stay-type visitors, (c) longer stay-type random walkers, and (d) shorter stay-type random walkers. The black dashed line shows the CCDF of the power law probability density function:

$$P(x) \propto x^{-\alpha} \qquad (3)$$

Our results show that the distribution of the path types for the shorter stay-type visitors as well as random walkers (red lines) can be approximated by power laws with exponents α=1.99 and 1.82) than the longer stay-type ones (blue lines, α=2.33 and 2.08, respectively). To test whether the power law is the best description for our dataset, we performed comparative analysis of the goodness of fit of another possible candidate distributions for our datasets, and exponential function. We compared the power law and exponential functions to our four individual datasets through the maximum likelihood method as well as the log likelihood ratio test provided by Alstott et al., (2014). The results indicate that all datasets except the longer stay-type visitors are better fit by the power laws than the exponential functions (p-value<0.01), and the longer stay-type random walkers fit better with lognormal than with the power law distribution (p-value=0.02).

We can interpret the obtained results as follows: Many shorter stay-type visitors tend to use the same paths from the entrance to the exit, so a few path types have a much higher frequency, while most path types appear only once. Conversely, most of the longer stay-type visitors also use a few path types, but their frequencies are not so high as those for the shorter stay-type visitors. Rather, most of the longer stay-type visitors are likely to follow different trajectories. As a result, the shorter visitors' lengths of stay become, the more the visitors follow a similar sequence in the museum.

### 3.2 Patterns of visitors' sequential movement by R-value

Table 3. Top five paths with the highest frequency, ordered by the observed paths in each group.

| Path | OP | RW | R | ρ | p-value |
|---|---|---|---|---|---|
| *Longer Stay Type Visitors* | | | | | |
| E-S-E | 0.216 | 0.0362 | 5.859 | 0.753 | 0.0117 |
| E-D-S-B-E | 0.058 | 0.0179 | 3.246 | 0.182 | 0.6140 |
| E-D-S-B-D-V-C-E | 0.028 | 0.0012 | 22.558 | -0.863 | 0.0012 |
| E-B-E | 0.024 | 0.0129 | 1.898 | 0.109 | 0.7634 |
| E-D-S-B-D-E | 0.023 | 0.0039 | 6.012 | -0.899 | 0.0003 |
| *Shorter Stay Type Visitors* | | | | | |
| E-S-E | 0.123 | 0.0455 | 2.703 | 0.753 | 0.0117 |
| E-D-S-B-D-V-C-E | 0.060 | 0.0045 | 13.392 | -0.863 | 0.0012 |
| E-D-S-B-E | 0.056 | 0.0316 | 1.780 | 0.182 | 0.6140 |
| E-D-S-B-D-E | 0.050 | 0.0098 | 5.120 | -0.899 | 0.0003 |
| E-D-S-B-D-V-E | 0.040 | 0.0023 | 17.349 | -0.522 | 0.1210 |

Table 3 presents the top five paths with the highest frequency, ordered by the observed paths in each group. Within them, E-S-E appears with the highest frequency in both groups. In addition, we can observe that its frequency is higher among the longer stay-type visitors than among the shorter stay-type visitors. More than 20% of the longer stay-type visitors visit only one node and explore other areas without passing through any other nodes. However, if we focus on R-value, the third-highest frequency's path, E-D-S-B-D-V-C-E, appears with the highest R-value among other paths among the longer stay-type visitors. This indicates that, although the former path (i.e., E-S-E) has a higher frequency than the latter (i.e., E-D-S-B-D-V-C-E), the latter has a stronger pattern than the former. In a similar way, E-B-S-D-V-C-E has a much higher R-value than other paths among the shorter stay-type visitors, although its frequency is lower ranked (eighth place).

**Table 4.** All paths that satisfy both the established thresholds (i.e., ρ>0.6%, p<0.01%).

| Path | R | ρ | p | Frequency (rank) |
|---|---|---|---|---|
| *Longer Stay Type Visitors* | | | | |
| E-D-S-B-S-D-E | 21.16 | 0.88 | 0.0006 | 0.007 (18) |
| E-C-V-C-E | 0.58 | 0.79 | 0.0062 | 0.002 (54) |
| *Shorter Stay Type Visitors* | | | | |
| E-B-S-D-V-C-E | 369.708 | -0.87 | 0.0009 | 0.02 (8) |
| E-B-S-D-V-E | 213.108 | -0.79 | 0.0065 | 0.007 (28) |
| E-D-S-V-B-D-E | 165.638 | -0.88 | 0.0005 | 0.003 (48) |
| E-D-S-B-D-B-V-E | 148.343 | -0.77 | 0.0084 | 0.0008 (113) |
| E-V-D-S-D-B-E | 89.680 | -0.78 | 0.0069 | 0.003 (51) |
| E-C-V-D-S-B-E | 71.176 | -0.91 | 0.0002 | 0.027 (7) |
| E-V-C-D-S-D-E | 67.285 | -0.87 | 0.0008 | 0.006 (33) |
| E-C-V-D-S-E | 45.376 | -0.83 | 0.0024 | 0.025 (10) |
| E-V-C-G-D-S-E | 22.365 | -0.84 | 0.002 | 0.0013 (81) |
| E-C-G-D-S-E | 14.276 | -0.85 | 0.0014 | 0.0013 (82) |
| E-D-S-B-D-V-C-E | 13.392 | -0.86 | 0.0012 | 0.06 (2) |
| E-S-D-V-C-E | 11.761 | -0.96 | 1.02E-05 | 0.03 (6) |
| E-V-D-S-B-E | 8.35 | -0.92 | 0.0001 | 0.018 (12) |
| E-S-D-V-E | 5.507 | -0.81 | 0.0041 | 0.008 (26) |
| E-D-S-B-D-E | 5.12 | -0.89 | 0.0003 | 0.05 (4) |
| E-D-S-V-C-E | 4.80 | -0.76 | 0.0097 | 0.02 (9) |
| E-D-V-C-D-S-B-E | 4.16 | -0.79 | 0.0059 | 0.001 (69) |
| E-D-V-C-S-E | 2.13 | -0.80 | 0.0046 | 0.0008 (114) |
| E-D-S-D-E | 0.74 | -0.80 | 0.0052 | 0.0048 (38) |

Table 4 presents all paths that satisfy both of the above-mentioned thresholds (i.e., ρ>0.6, p<0.01). We sort them by the R-value. Nineteen paths are extracted from the shorter stay-type visitors, but only two paths are discovered from the longer stay-type visitors. Here again, a larger number of path types appears among the shorter stay-type visitors than among the longer stay-type visitors. In addition, the R-value of the shorter stay-type visitors is significantly larger than that of the longer stay-type visitors. For example, the strongest R-value from the longer stay-type visitors is similar to the one ranked 10th in the shorter stay-type visitors, and the second strongest one (i.e., E-C-V-C-E) is lower than all R-values from the longer stay-type visitors. Actually, the strongest R-value from the shorter stay-type visitors (i.e., E-B-S-D-V-C-E) is 17 times larger than the strongest R-value from the longer stay-type visitors (i.e., E-D-S-B-S-D-E).

All these facts indicate that the most frequently appearing pattern from the shorter stay-type visitors is quite strong, and that of the longer stay-type visitors is relatively weak. This is probably because the shorter stay-type visitors have a limited available time, so they might follow optimized paths to be able to visit the "must-see" art pieces in the most efficient way. As a result, the visiting style of the longer stay-type visitors shows higher diversity than that of the shorter stay-type. The strength of this selectivity gets weaker when the length of stay increases.

## 4. Discussion

Previous research discovered E-D-S-B-D-V-C-E as the most frequently appearing path from both groups (Yoshimura et al., 2014), but the current research specifies this path as the determinant factor for the shorter stay-type visitors. This is because Spearman's rank correlation between the frequency of visitors' paths and their lengths of stay in the museum is negative, meaning that the distribution of E-D-S-B-D-V-C-E is unique to the shorter stay-type ($\rho=-0.86$, $p=0.0012$). In addition, we find that most R-values (i.e., observed frequency/random walk frequency) from the shorter stay-type visitors are much larger than those from the longer stay-type visitors. For example, the average R-value of the shorter stay-type is 3.7 times bigger than the longer stay-type. This suggests that the extracted patterns from the shorter stay-type visitors are much stronger than the ones from the longer stay-type visitors.

All these facts indicate that the shorter stay-type visitors show much stronger mobility patterns than the longer stay-type visitors, and this coincides with our intuition. We intuitively consider that the shorter stay-type visitors would follow more similar paths than the longer stay-type visitors because they might want to follow the optimized paths that enable them to visit all the "must-see" art pieces within their limited available time. There are very few such paths given the structure of the museum's network. As a consequence, the average number of visited locations by the shorter stay-type visitors becomes frequently similar to that of the longer stay-type visitors, although the difference of each group's length of stay in the museum is more than 4 hours. All of our presented analysis validates this aspect using our empirically collected large-scale dataset.

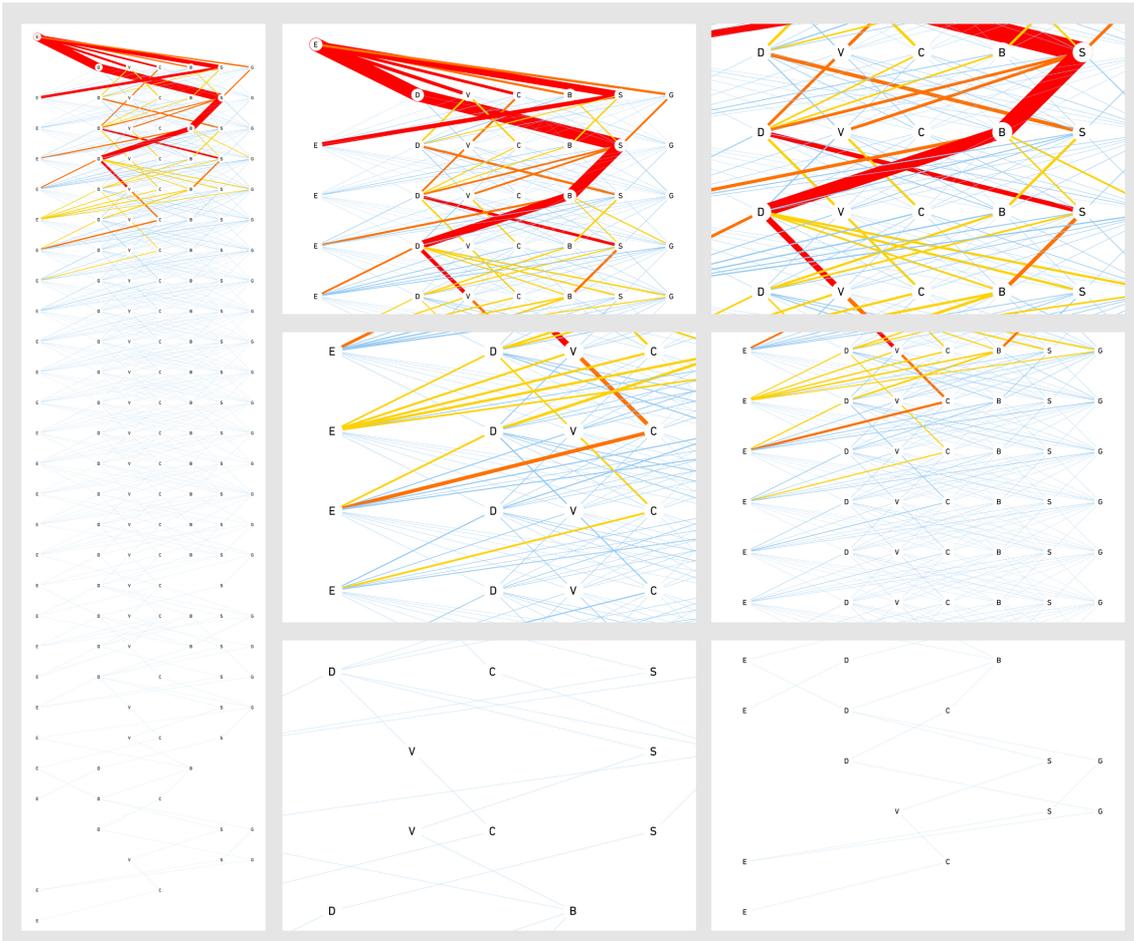

**Figure 5.** The most frequently appearing paths, which emerged in a bottom up way

We visualize all these results in Figure 5, in which we trace all visitors' sequential movements and identify the most used ones in a bottom-up way. We can see some of the frequently appearing paths until the path sequence length is 8, but such bold lines disappear when the path sequence length becomes more than 8.

Our approach is opposite to the conventional one, which largely depends on small samples of visitors' behaviors derived from interviews or questionnaires (Yalowitz and Bronnenkant, 2009). Furthermore, our analytical framework and applied technique present an alternative to the application of sequence alignment methods to human movement datasets such as those of tourists' movements in urban settings (Shoval and Isaacson, 2007; Shoval et al., 2013) and pedestrians' movements in a city (Versichele, et al., 2011) and inside a building (Delafontaine et al., 2012; Kanda et al., 2007). Although the sequence alignment method can measure the relative distance of the path in comparison with all other sequential paths by their similarities, the result depends greatly on the weight and parameters given in advance, resulting in the questionable validity and reliability of this technique (Shoval et al., 2013).

As a backdrop to these situations, the proposed methodology can make a reference line that enables us to measure and evaluate the distance of the observed paths. Since the random walker randomly chooses a path with the same probability, a shorter or longer path sequence appears independently of its path sequence length.

As a consequence, this methodology enables us to compare the frequency of different path sequence lengths in the same framework.

In this way, our approach fills in the gaps and drawbacks of the most of the current research in art museums and galleries. The empirically accumulated dataset enables us to shed light on unknown aspects of visitors' behaviors in a dynamic way. This is opposite to a conventionally conducted analysis, which uses small samples of visitors' behaviors. The obtained results and findings are useful for managing spatial congestion through controlling visitor flow but also risk mapping. In addition, we can optimize the museum's facilities by considering visitors' routes and lengths of stay at each specific artwork as well as in the museum. Finally, our analysis and results provide significant information with which to create or improve interactive applications for mobile devices such as audio guides and visit guidance at large.

**Appendix (or "Material and Methods")**

Algorithm for sequences extraction. Let's define $S_i = \{s_{1i}, s_{2i}... s_{pi} ... s_{Pi}\}$ the set of all the $P$ possible sequences of length $i$, where $i = \{k, k+1, k+2...I\}$, where I is the length of the longest trajectory in the dataset, and k is the minimum meaningful trajectory in the dataset. Let's also define $f(s_{pi})$ as the number of visitors in the dataset that used the sequence $s_{pi}$ during their visit to the museum. Finally, lets define T as a table containing the resulting patterns and their frequencies.

Given that $N = \{1,2...n\}$, the set of nodes in the museum, the high-level steps of the algorithm are as follow:

```
1     for i in k to I:
2         for s_pi in S_i:
3             if f(s_pi) > 0
4                 save s_pi in T
5                 save f(s_pi) in T
6                 for n in Neighbors(s_pi[i]):
7                     add s_{pi+1}^n to S_{i+1}
```

being that $s_{pi+1}^n$ is a sequence of length $i+1$ that results from adding a node $n$ belonging to the neighbors of the last node of the sequence $s_{pi}$ (i.e. $s_{pi}.[i]$), to the sequence $s_{pi}$.

This algorithm starts from the basis that any sequence including a subsequence cannot be present if such subsequence is not present in the dataset. It then iteratively finds all sequences of a minimum length $k$ that are actually used by at least one visitor. Then builds all the possible sequences of length k+1 based on the existing sequences of length k and discarding the inexistent ones. By discarding the shortest-length sequences, the algorithm converges faster than if every possible sequence was tested.

We used $k = 3$ as a starting sequence length, based on the fact that the shortest possible trajectory, (e.g. 0-8-0 or 0-3-0) has a length of three. The resulting table T includes all sequences that appear in the dataset, along with their respective frequencies. Not every possible sequence is going to appears in T, since there are paths that are impossible for visitors to follow due to the physical distribution of the museum.

**Spearman's correlation coefficient.** The Spearman's correlation coefficient ρ (Corder & Foreman, 2009) is a non-parametric measure of statistical dependence between two variables. The coefficient evaluates how well the relationship between two variables (x, y) can be described by a monotonic function. The coefficient assumes values between -1 (where $\frac{dy}{dx} < 0 \; x \in \mathbb{R}$) and +1 (where $\frac{dy}{dx} > 0 \; x \in \mathbb{R}$), the extremes reached when one of the variable is a perfect monotone of the other. A correlation coefficient of zero indicates there is no tendency for *y* to increase or decrease as *x* increases. So, if x and y are the variables to correlate, *xi* and *yi* are the ranked values, then the Spearman's correlation coefficient can be calculated from:

$$\rho = \frac{\sum_i (x_i - \bar{x})(y_i - \bar{y})}{\sqrt{\sum_i (x_i - \bar{x})^2 \sum_i (y_i - \bar{y})^2}}$$

## 6. Reference

Bourdeau L, Chebat J C, 2001, "An Empirical Study of the Effects of the Design of the Display Galleries of an Art Gallery on the Movement of Visitors" *Museum Management and Curatorship* **19**(1) 63-73.
Boccaletti S, Latora V, Moreno Y, Chavez M, Hwang D U, 2006, "Complex networks: Structure and dybamics" *Physics Reports* 424 175-308
Corder G W & Foreman D I, 2009, Nonparametric statistics for non-statisticians: A step-by-step approach. Hoboken, N.J: Wiley.
Delafontaine M, Versichele M, Neutens T, Van de Weghe N, 2012, "Analysing spatiotemporal sequences in Bluetooth tracking data" *Applied Geography* **34** 659-668
Eagle N, Pentland A, 2005, "Reality mining: sensing complex social systems" *Personal and Ubiquitous Computing* **10**(4) 255-268
Hein G, 1998 *Learning in the Museum* (Routledge, London)
Hooper-Greehhill E, 1995, "Audiences- A Curatorial Dilemma", in *Art in Museums* Ed Pearce S (London & Atlantic Highlands, NJ) pp 143-163
Hooper-Greenhill E, 2006, "Studying visitors", in *A Companion to Museum Studies* Ed MacDonald S (Blackwell Publishing, London) pp362-376
Kanda T, Shiomi M, Perrin L, Nomura T, Ishiguro H, Hagita N, 2007, "Analysis of people trajectories with ubiquitous sensors in a science museum" *Proceedings 2007 IEEE International Conference on Robotics and Automation (ICRA'07)* 4846-4853
Kostakos V, O'Neill E, Penn A, Roussos G, Papadongonas D, 2010, "Brief encounters: sensing, modelling and visualizing urban mobility and copresence networks" *ACM Transactions on Computer Human Interaction* **17**(1) 1-38
Levasseur M, Veron E, 1993, "Ethnographie de l'exposition. L'espace, le corps, le sens", Paris: BPI du Centre Pompidou, 179p.
Loomis R J, 1987 *Museum Visitor Evaluation: New Tool for Management* (American Association for State and Local History, Nashville, TN)
Mayer-Schönberger V, Cukier K, 2013, *Big Data: A Revolution That Will Transform How We Live, Work and Think* (John Murray, London)
Melton A W, 1935, *Problems of Installation in Museums of Art. American Association of Museums Monograph New Series No. 14* (American Association of Museums, Washington, DC)
Nielsen L C, 1942, "A Technique for Studying the Behavior of Museum Visitors", *The Journal of Educational Psychology* **37** 103-10
Paulos E, Goodman E, 2004, "The familiar stranger: anxiety, comfort, and play in public places" *Proceedings of the SIGCHI conference on Human factors in computing systems* **6**(1) 223-230.
Robinson E S, 1928, *The Behavior of the Museum Visitor. American Association of Museums Monograph New Series No. 5* (American Association of Museums, Washington, DC)
Sanfeliu A, Llácer M R, Gramunt M D, Punsola A, Yoshimura Y, 2010, "Influence of the privacy issue in the Deployment and Design of Networking Robots in European Urban Areas" *Advanced Robotics* **24**(13) 1873-1899
Serrell B, 1998 *Paying Attention: Visitors and Museum Exhibitions* (American Association of Museums, Washington DC)

Shoval N, McKercher B, Birenboim A, Ng E, 2013, "The application of a sequence alignment method to the creation of typologies of tourist activity in time and space" *Environment and Planning B: Planning and Design advance online publication*, doi:10.1068/b38065

Shoval, N., & Issacson, M. (2006). Application of tracking technologies in the study of pedestrian spatial behavior, The Professional Geographer, 58, 172-183.

Sinatra R, Condorelli D, Latora V, 2010, "Networks of motifs from sequences of symbols" *Physical review letters* 105 (17), 178702.

Sinatra R, Gómez-Gardenes J, Lambiotte R, Nicosia V, Latora V, 2011, "Maximal-entropy random walks in complex networks with limited information" *Physical Review E* 83(3), 030103.

Sparacino F, 2002, "The Museum Wearable: real-time sensor-driven understanding of visitors's interests for personalizad visually-augmented museum experiences" in *Museums and the Web 2002: Proceedings* Eds Bearman D, Trant J (Archives & Museum Informatics, Tronto)

Stallings W, 2011, *Cryptography and Network Security: Principles and Practice, 5th Edition* (Prentice Hall, Boston MA)

Tschacher W, Greenwood S, Kirchberg V, Wintzerith S, Van den Berg Karen, Tröndle M, 2012, "Physiological correlates of aesthetic perception of artworks in a museum" *Psychology of Aesthetics, Creativity, and the Art* **6** (1) 96-103

Versichele M, Neutens T, Delafontaine M, Van de Weghe N, 2011, "The use of Bluetooth for analysing spatiotemporal dynamics of human movement at mass events: a case study of the Ghent festivities" *Applied Geography* **32** 208-220

Webb E J, Campbell D T, Schwartz R D, Sechrest L (2000) Unobtrusive measures: revised edition. Thousand Oaks: Saga Publications, New York

Wittlin A S, 1949, *The Museum, Its History and its Tasks in Education*, (Routledge and Kegan Paul, London)

Yalowitz S S, Bronnenkant Kerry, 2009, "Timing and Tracking: Unlocking Visitor Behavior" *Visitor Studies* **12**(1) 47-64

Yoshimura Y, Girardin F, Carrascal J P, Ratti C, Blat J, 2012, "New Tools for Studing Visitor Behaviours in Museums: A Case Study at the Louvre" in *Information and Communication Technologis in Tourism 2012. Proceedings of the International conference in Helsingborg (ENTER 2012)* Eds Fucks M, Ricci F, Cantoni L (Springer Wien New York, Mörlenback) 391-402

Yoshimura Y, Krebs A, Ratti C, 2017, "Noninvasive Bluetooth Monitoring of Visitors' Length of Stay at the Louvre" IEEE Pervasive Computing 16 (2), p26-34.

Yoshimura Y, Sobolevsky S, Ratti C, Girardin F, Carrascal J P, Blat J, Sinatra R, 2014, "An analysis of visitors' behaviour in The Louvre Museum: a study using Bluetooth data" Environment and Planning B: Planning and Design 41 (6) 1113-1131